\documentstyle[preprint,pra,aps]{revtex}
\begin{document}

\draft

\title{Atomic Bose gas with negative scattering length}

\author{H.T.C. Stoof}
\address{University of Utrecht, Institute for Theoretical
         Physics, Princetonplein 5, \\
         P.O. Box 80.006, 3508 TA  Utrecht, The Netherlands}

\maketitle

\begin{abstract}
We derive the equation of state of a dilute atomic Bose gas with an interatomic
interaction that has a negative scattering length and argue that two continuous
phase transitions, occuring in the gas due to quantum degeneracy effects, are
preempted by a first-order gas-liquid or gas-solid transition depending on the
details of the interaction potential. We also discuss the consequences of this
result for future experiments with magnetically trapped spin-polarized atomic
gasses such as lithium and cesium.
\end{abstract}


\section{INTRODUCTION}
Apart from the intrinsic interest in cooling and trapping atomic gas samples by
means of electromagnetic fields \cite{chu} and their potential use in various
high-precision experiments, the possibility of achieving Bose-Einstein
condensation is also an important motivation for studying ultracold atomic
gasses in a static magnetic trap. The first steps towards this goal were made
with atomic hydrogen \cite{mit,adam} and recent progress in lowering the
temperature of the trapped gas using both conventional \cite{doyle,luiten} and
light-induced \cite{setija} evaporative cooling shows that atomic hydrogen is
still a very promising candidate for the ultimate achievement of Bose-Einstein
condensation.

Due to their excellent optical properties a substantial experimental effort is
presently also directed to the alkali-metal vapors cesium \cite{wieman} and
lithium \cite{hulet}. However, to ensure the stability of the condensed phase
it is usually (but see below) required that the interaction between the atoms
is effectively repulsive or more precisely that its scattering length $a$ is
positive \cite{fetter}. There is no doubt that the above requirement is
fulfilled for atomic hydrogen, but in the case of atomic lithium the
state-of-the-art triplet potential leads to a negative scattering length for
the doubly spin-polarized $|F=2, M_F=2 \rangle$ state \cite{ard}. In contrast,
the situation for atomic cesium is less straightforward. It has recently been
shown that with the present knowledge of the interaction potentials it is not
possible to determine the sign of $a$ for the doubly spin-polarized $|F=4,
M_F=4 \rangle$ state, but that the scattering length for the also low-field
seeking $|F=3, M_F=-3 \rangle$ state of the lower hyperfine manifold has a
pronounced resonance structure, which offers the exciting possibility of
controlling the sign of $a$ by an appropriate choice of the magnetic-field
strength \cite{eite}.

In view of these circumstances it is of considerable interest to investigate
the properties of a dilute atomic Bose gas with negative scattering length and
to predict what one might observe in experiments with magnetically trapped
atomic lithium and cesium. In particular, it is interesting to see if also for
these gasses quantum degeneracy leads to Bose-Einstein condensation at
sufficiently high densities. However, before we can begin with a detailed
discussion of these issues we must be somewhat more precise and realize that
besides the scattering length $a$, an interatomic potential is also
characterized by the hard-core radius $r_C$, the finite range $r_V$ and the
well-depth $\epsilon_W$. In terms of these quantities the main difference
between spin-polarized atomic hydrogen and spin-polarized atomic lithium and
cesium is that in the former case the potential well is so shallow that it
admits no bound states, whereas in the latter case $\epsilon_W \gg
\hbar^2/mr_V^2$ and many rovibrational states are possible.

As a result the relationship between the pressure $p$ and the inverse density
$1/n$ at a fixed and nonzero temperature
$T \ll \epsilon_W/k_B$ is qualitatively different and as shown schematically in
Fig.\ \ref{fig1} \cite{foot}. At these temperatures atomic lithium and cesium
may thus undergo a first-order transition from a dilute ($nr_V^3 \ll 1$)
gaseous phase to a high density ($nr_C^3 = O(1)$) phase, which is either liquid
or solid depending on the details of the interaction potential. The coexistence
line of these two phases is obtained from a Maxwell construction and is
therefore such that the areas of region I and II are equal \cite{huang}. Note
that the critical pressure $p_c(T)$ found in this manner is always larger than
zero, because the area of region I is bounded and the area of region II can be
made arbitrarily large since for low densities ($n\Lambda^3 \ll 1$ where
$\Lambda = \sqrt{2 \pi \hbar^2/mk_BT}$ is the thermal de Broglie wavelength)
the equation of state reduces to $p=nk_BT$. Hence, for sufficiently low
densities the gaseous phase is stable and we are allowed to incorporate the
influence of the interaction by means of perturbation theory around the ideal
Bose gas. Moreover, at the experimental densities of interest, we are justified
in using the T-matrix or ladder approximation because $nr_V^3 \ll 1$ and we
only need to include all two-body processes in our description of the dilute
phase.

For an adequate determination of the coexistence line, however, a more advanced
theory is needed which is capable of describing a strongly interacting system
at high densities. Fortunately, in the case of magnetically trapped atomic
gasses we are dealing with a system that is expected to have a large
free-energy barrier for the formation of a critical bubble of the dense phase.
Hence, if the density is isothermally increased beyond the point of
coexistence, a metastable supersaturated vapor will be formed. From an
experimental point of view the relevant question to be answered is therefore:
Can Bose-Einstein condensation also occur in the (meta)stable region of the
phase diagram (where $\partial p/\partial n >0$) of an atomic Bose gas with
negative scattering length? The answer to this question can still be found
within the framework of the ladder approximation and is the main topic of the
present paper.

In order to do so we must consider the degenerate regime where $n\Lambda^3 =
O(1)$ and allow for the possibility of a second-order phase transition due to
Bose degeneracy. Consequently, we must first of all identify the appropriate
order parameter. This is achieved in Sec.\ \ref{order}. Although Sec.\
\ref{order} is certainly an important part of the paper, the discussion
presented there is of a somewhat technical nature and makes use of functional
methods which are very convenient for the  derivation of the free-energy
density of the order parameters of interest. However, after the correct order
parameter is found we return in Sec.\ \ref{stable} to canonical methods in
order to arrive at the final results of the paper in a physically more
transparent, and perhaps more accessible manner. In particular, we derive in
Sec.\ \ref{stable} the equation of state of the gas and show that it indeed, in
the region of validity of the T-matrix approximation, leads to a relationship
between the pressure and the inverse density which is qualitatively in
agreement with Fig.\ \ref{fig1}. Most importantly, we are then in a position to
answer the question mentioned above. We end in Sec.\ \ref{concl} with some
conclusions and with a discussion of the experimental relevance of our results.

\section{ORDER PARAMETER}
\label{order}
We consider a homogeneous system of $N$ bosonic atoms in a volume $V$. Using a
functional approach to the imaginary time formalism, the Euclidean action of
this system is
\begin{eqnarray}
\label{action}
S[\psi^*,\psi] &=& \int_{0}^{\hbar \beta} d\tau
             \int d\vec{x} \: \psi^*(\vec{x},\tau)
             \left( \hbar \frac{\partial}{\partial \tau} -
                    \frac{\hbar^2 \nabla^2}{2m} -\mu \right)
                                  \psi(\vec{x},\tau) \nonumber \\
          &+& \: \frac{1}{2} \int_{0}^{\hbar \beta} d\tau
                                  \int d\vec{x} \int d\vec{x}' \:
                 \psi^*(\vec{x},\tau) \psi^*(\vec{x}',\tau)
                    V(\vec{x}-\vec{x}')
                      \psi(\vec{x}',\tau) \psi(\vec{x},\tau) \:,
\end{eqnarray}
where $\beta = 1/k_BT$, $\mu$ is the chemical potential, $V(\vec{x}-\vec{x}')$
is the interatomic interaction potential and possible three-body forces are
neglected since we are interested in the dilute limit. In addition, the grand
canonical partition function is then given by the functional integral
\begin{equation}
\label{z}
Z_G = \int d[\psi^*] d[\psi] \:
       exp \left( - \frac{1}{\hbar} S[\psi^*,\psi] \right)
\end{equation}
over the periodic c-number fields $\psi^*(\vec{x},\tau)$ and
$\psi(\vec{x},\tau)$. For the ultracold gasses considered here the range of the
interaction is always much smaller than the thermal de Broglie wavelength
($r_V/\Lambda \ll 1$), which implies that we can replace the potential
$V(\vec{x}-\vec{x}')$ by the contact interaction $V_0 \delta(\vec{x}-\vec{x}')$
with $V_0 \equiv \int d\vec{x} \: V(\vec{x})$. Together with the thermodynamic
potential $\Omega = - (lnZ_G)/\beta$ this completely determines the
thermodynamic properties of the weakly-interacting Bose gas.

\subsection{Bose-Einstein condensation}
Nevertheless, to extract useful information from Eqs.\ (\ref{action}) and
(\ref{z}) we must resort to perturbation theory and thus incorporate the
correct symmetry of the homogeneous and at high densities metastable state of
the gas. Clearly, both objectives can be achieved if we are able to identify
the correct order parameter. An immediate choice is the thermal average
$\langle \psi(\vec{x},\tau) \rangle$. However, this choice turns out to be
inappropriate for the atomic Bose gas with negative scattering length as can
already be seen from the following {\it ad absurdum} argument: Assuming
$\langle \psi(\vec{x},\tau) \rangle$ to be non zero, the application of the
usual Bogoliubov theory shows that the long-wavelength fluctuations in the
order parameter are unstable for $a<0$ \cite{fetter}, which invalidates the
initial assumption.

A more rigorous argument is based on the free-energy functional for the time
and space-independent value of the above order parameter
\begin{equation}
F[\langle \psi^* \rangle, \langle \psi \rangle]
   = V \sum_{n=1}^{\infty}
            \frac{\Gamma_0^{(2n)}}{(n!)^2}
                   |\langle \psi \rangle|^{2n} \:,
\end{equation}
where $\Gamma_0^{(2n)}$ is the $2n$-point vertex function with all $2n-1$
momentum and frequency arguments equal to zero \cite{amit,stoof}. Using the
T-matrix approximation the latter can easily be evaluated in the symmetric (low
density or high temperature) phase with the result \cite{stoof}
\begin{mathletters}
\label{vertex}
\begin{equation}
\Gamma_0^{(2)} = \hbar \Sigma(\vec{0};0) - \mu \equiv - \mu' \:,
\end{equation}
\begin{equation}
\label{g4}
\Gamma_0^{(4)} = 2T^{MB}(\vec{0},\vec{0},\vec{0};2\mu') \:,\:
                                                      {\rm and}
\end{equation}
\begin{equation}
\Gamma_0^{(2n)} = 0 \:,\: {\rm for \:} n \geq 3.
\end{equation}
\end{mathletters}

\noindent
In deriving Eq.\ (\ref{g4}) we also made use of the fact that for the momenta
$\hbar k \ll \hbar/r_V$ and Matsubara frequencies $\omega_n \ll \hbar/mr_V^2$
of interest the selfenergy $\hbar \Sigma(\vec{k};\omega_n)$ is well
approximated by the constant $2nT^{2B}(\vec{0},\vec{0};0) = 8\pi na \hbar^2/m$.
The selfenergy thus only leads to a shift in the chemical potential and we find
that $\mu'$ is obtained from the equation of state of the ideal Bose gas
\begin{equation}
\label{n}
n = \frac{1}{V} \sum_{\vec{k}} N(\epsilon_{\vec{k}}-\mu') \:,
\end{equation}
with $\epsilon_{\vec{k}} = \hbar^2 \vec{k}^2/2m $ the kinetic energy of the
atoms and $N(\epsilon)$ the Bose distribution $1/(e^{\beta \epsilon}-1)$. In
addition, the many-body T matrix can now be expressed in terms of the two-body
T matrix $T^{2B}(\vec{0},\vec{0};0) = 4\pi a \hbar^2/m$ via \cite{stoof}
\begin{equation}
\label{tmb}
\frac{1}{T^{MB}(\vec{0},\vec{0},\vec{0};2\mu')}
 = \frac{1}{T^{2B}(\vec{0},\vec{0};0)}
 + \frac{1}{V} \sum_{\vec{k}}
   \frac{N(\epsilon_{\vec{k}}-\mu')}{\epsilon_{\vec{k}}-\mu'} \:.
\end{equation}

Through Eqs.\ (\ref{vertex}), (\ref{n}) and (\ref{tmb}) we observe that for an
effectively repulsive interaction $(a>0)$ the free-energy functional has the
same form as in the Landau theory of second-order phase transitions \cite{ma}.
In particular, the free energy is bounded from below and acquires at
temperatures below the critical temperature of the ideal Bose gas
\begin{equation}
T_0 = \frac{2\pi \hbar^2}{mk_B}
         \left( \frac{n}{\zeta(\frac{3}{2})} \right)^{2/3}
\end{equation}
a minimum for $\langle \psi \rangle \equiv \sqrt{n_0}e^{i\vartheta}$, which
determines the condensate density $n_0$. However, for an effectively attractive
interaction $(a<0)$ this is not true, because the coefficient of the quartic
term in the free-energy density is negative. We are therefore again led to the
conclusion that in the case of an atomic Bose gas with negative scattering
length $\langle \psi(\vec{x},\tau) \rangle$ is not the correct order parameter
and that Bose-Einstein condensation in this canonical sence does not take
place.

\subsection{Evans-Rashid transition}
Instead, an analogy with the BCS theory of superconductivity suggests that if
the atoms have an effectively attractive interaction, a phase transition with
the order parameter $\Delta(\vec{x},\tau) \equiv V_0 \langle \psi(\vec{x},\tau)
\psi(\vec{x},\tau) \rangle$ occurs. To see if this is indeed the case we must
derive and solve the equation for the equilibrium value of
$\Delta(\vec{x},\tau)$ in the ladder approximation, which requires the
inclusion of fluctuations around the usual BCS (mean-field) theory. As a first
step towards this goal we perform a Hubbard-Stratonovich transformation by
multiplying the grand canonical partition $Z_G$ by
\begin{equation}
1 = {\cal N} \int d[\Delta^*] d[\Delta] \:
      exp \left( \frac{1}{2\hbar V_0}
                   \int_0^{\hbar \beta} d\tau \int d\vec{x} \:
                   |\Delta(\vec{x},\tau) -
                     V_0 \psi(\vec{x},\tau) \psi(\vec{x},\tau)|^2
                              \right)
\end{equation}
and integrating over the fields $\psi^*(\vec{x},\tau)$ and
$\psi(\vec{x},\tau)$. The latter is conveniently accomplished in Nambu space
and requires the introduction of the two-component field $\phi(\vec{x},\tau)
\equiv ( \psi(\vec{x},\tau), \psi^*(\vec{x},\tau) )$ and the corresponding
matrix of exact (normal and anomalous) one-particle Green's functions
$G(\vec{x},\tau;\vec{x}',\tau') \equiv
      - \langle T_{\tau}[\phi(\vec{x},\tau)
                              \phi^{\dagger}(\vec{x}',\tau')]
                                                      \rangle $,
obeying the Gorkov equation
\begin{equation}
\left(
  \begin{array}{cc}
      - \hbar \frac{\partial}{\partial \tau}
         + \frac{\hbar^2 \nabla^2}{2m} + \mu &
                                       - \Delta(\vec{x},\tau) \\
      - \Delta^*(\vec{x},\tau) &
             \hbar \frac{\partial}{\partial \tau}
         + \frac{\hbar^2 \nabla^2}{2m} + \mu \\
  \end{array} \right)
G(\vec{x},\tau;\vec{x}',\tau') = \hbar \delta(\vec{x}-\vec{x}')
                                           \delta(\tau-\tau') \:.
\end{equation}

In this manner we arrive at an effective action for the order parameter
$\Delta(\vec{x},\tau)$, which is formally given by \cite{kleinert}
\begin{equation}
S[\Delta^*,\Delta] = \frac{\hbar}{2} Tr[ ln(-G^{-1}) ]
       - \frac{1}{2V_0}
            \int_0^{\hbar \beta} d\tau \int d\vec{x} \:
                     |\Delta(\vec{x},\tau)|^2
\end{equation}
and contains all the desired information on the possibility of a BCS-like phase
transition. In particular, it can be expanded in powers of
$\Delta(\vec{x},\tau)$ by using the Dyson equation $G^{-1} = G_0^{-1} -
\Sigma$, leading to
\begin{equation}
Tr[ ln(-G^{-1}) ] = Tr[ ln(-G_0^{-1}) ]
      - \sum_{n=1}^{\infty} \frac{Tr[(G_0 \Sigma)^n]}{n} \:,
\end{equation}
and by taking the selfenergy matrix equal to
\begin{equation}
\hbar \Sigma(\vec{x},\tau;\vec{x}',\tau') =
   \left(
   \begin{array}{cc}
      0 & \Delta(\vec{x},\tau) \\ \Delta^*(\vec{x},\tau) & 0 \\
   \end{array} \right)
                \delta(\vec{x}-\vec{x}') \delta(\tau-\tau') \:.
\end{equation}
For our purposes we are especially interested in the quadratic term in this
expansion. After performing the trace over both coordinate and Nambu space it
is found to be
\begin{equation}
S^{(2)}[\Delta^*,\Delta] =
  - \int_0^{\hbar \beta} d\tau \int d\vec{x}
     \int_0^{\hbar \beta} d\tau' \int d\vec{x}' \:
        \Delta^*(\vec{x},\tau)
           \hbar G_{\Delta}^{-1}(\vec{x},\tau;\vec{x}',\tau')
                          \Delta(\vec{x}',\tau') \:,
\end{equation}
where the `noninteracting' Green's function of the order parameter obeys
\begin{equation}
G_{\Delta}^{-1}(\vec{x},\tau;\vec{x}',\tau') =
   \frac{1}{2\hbar^2} G_{0,11}(\vec{x},\tau;\vec{x}',\tau')
                   G_{0,11}(\vec{x},\tau;\vec{x}',\tau')
 + \frac{1}{2\hbar V_0} \delta(\vec{x}-\vec{x}') \delta(\tau-\tau')
\end{equation}
or equivalently
\begin{eqnarray}
G_{\Delta}(\vec{x},\tau;\vec{x}',\tau') &=&
    2\hbar V_0 \delta(\vec{x}-\vec{x}') \delta(\tau-\tau') \\
                                                    \nonumber
   &-& \frac{V_0}{\hbar}
          \int_0^{\hbar \beta} d\tau'' \int d\vec{x}'' \:
             G_{0,11}(\vec{x},\tau;\vec{x}'',\tau'')
             G_{0,11}(\vec{x},\tau;\vec{x}'',\tau'')
                 G_{\Delta}(\vec{x}'',\tau'';\vec{x}',\tau') \:.
\end{eqnarray}
Transforming to frequency-momentum space then gives
\begin{equation}
G_{\Delta}(\vec{K};\Omega_n) = 2\hbar V_0
   + \frac{1}{V} \sum_{\vec{k}} V_0
        \frac{ 1 + N(\epsilon_{\vec{K}/2+\vec{k}} - \mu)
                 + N(\epsilon_{\vec{K}/2-\vec{k}} - \mu) }
             { i\hbar \Omega_n + 2\mu
                         - \epsilon_{\vec{K}/2+\vec{k}}
                         - \epsilon_{\vec{K}/2-\vec{k}} }
                               G_{\Delta}(\vec{K};\Omega_n) \:,
\end{equation}
having the solution
\begin{equation}
\label{gdelta}
G_{\Delta}(\vec{K};\Omega_n) =
   2\hbar T^{MB}(\vec{0},\vec{0},\vec{K};i\hbar \Omega_n + 2\mu)
                                                            \:.
\end{equation}

In mean-field theory we neglect all fluctuations and as a result conclude that
the Evans-Rashid transition \cite{kleinert,evans} occurs at a temperature
determined by
\begin{equation}
\int_0^{\hbar \beta} d\tau \int d\vec{x} \:
  G_{\Delta}(\vec{x},\tau;\vec{x},\tau) =
     G_{\Delta}(\vec{0};0) =
       \frac{1}{2\hbar T^{MB}(\vec{0},\vec{0},\vec{0};2\mu)} = 0
                                                             \:,
\end{equation}
which corresponds exactly to the Thouless criterium for the onset of the BCS
instability \cite{thou}. For a dilute gas, however, mean-field theory is not
sufficiently accurate and we must also consider the fluctuations. Fortunately,
Eq.\ (\ref{gdelta}) shows how we can apply the T-matrix approximation to the
Evans-Rashid transition. Indeed, due to this relationship the renormalization
of the quadratic term in the action caused by the presence of the $|\Delta|^4$
and the $|\Delta|^6$ `interactions', can be accounted for by replacing in the
above discussion the noninteracting Green's function
$G_{0,11}(\vec{k};\omega_n) = \hbar/(i\hbar \omega_n - \epsilon_{\vec{k}} +
\mu)$ by its renormalized (within the ladder approximation) value
$\hbar/(i\hbar \omega_n - \epsilon_{\vec{k}} - \hbar \Sigma(\vec{k};\omega_n) +
\mu)$. Since $\hbar \Sigma(\vec{k};\omega_n)$ is well approximated by the
constant $2nT^{2B}(\vec{0},\vec{0};0)$ this implies just a replacement of the
chemical potential $\mu$ by $\mu'$. The onset of the BCS-like instability is
therefore in the dilute limit determined by
\begin{equation}
\label{ter}
\frac{1}{T^{MB}(\vec{0},\vec{0},\vec{0};2\mu')}
 = \frac{1}{T^{2B}(\vec{0},\vec{0};0)}
 + \frac{1}{V} \sum_{\vec{k}}
   \frac{N(\epsilon_{\vec{k}}-\mu')}{\epsilon_{\vec{k}}-\mu'} = 0
                                                              \:.
\end{equation}
Consequently, the critical temperature of the Evans-Rashid transition is
slightly above the critical temperature of the ideal Bose gas, i.e. $T_{ER} =
T_0(1-O(a/\Lambda_0))$ \cite{stoof}.

Notice that the above discussion shows that the pair field $\psi(\vec{x},\tau)
\psi(\vec{x},\tau)$ can be used even if the interatomic potential has a
negative scattering length without having a bound state. This is a truly
many-body effect which occurs because bosons prefer to scatter into (momentum)
states that are already occupied. As a result the binding between two particles
is effectively increased and a resonance, which must always lie just above the
continuum treshold for $a$ to be negative, becomes bound at a
density-temperature combination determined by Eq.\ (\ref{ter}).

\section{STABILITY OF THE GASEOUS PHASES}
\label{stable}
In the previous section we have argued that in the degenerate regime the
relevant order parameter for an atomic Bose gas with negative scattering length
is $\Delta(\vec{x},\tau) = V_0 \langle \psi(\vec{x},\tau) \psi(\vec{x},\tau)
\rangle$. In order to proceed and to discuss whether the corresponding phase
transition can take place in the (meta)stable region of the phase diagram we
must now also consider the gas below the critical temperature $T_{ER}$. This
can of course be achieved by the functional approach used above, but to make
the following more transparent we will from now on use canonical methods.

Denoting the equilibrium value of $\Delta(\vec{x},\tau)$ by $\Delta_0$ and
applying the BCS approach in combination with the T-matrix approximation, the
Hamiltonian of the gas is  approximated by
\begin{eqnarray}
\label{hamil}
H = \int d\vec{x} \:
    \left\{ \psi^{\dagger}(\vec{x})
           \left( - \frac{\hbar^2 \nabla^2}{2m} - \mu' \right)
           \psi(\vec{x})
           + \frac{\Delta_0}{2} \psi^{\dagger}(\vec{x})
                                   \psi^{\dagger}(\vec{x})
           + \frac{\Delta_0^*}{2} \psi(\vec{x}) \psi(\vec{x})
    \right\}         \nonumber \\
  - \frac{|\Delta_0|^2}{2V_0} - n^2 T^{2B}(\vec{0},\vec{0};0) \:,
\end{eqnarray}
in the Schr\"{o}dinger picture. After a diagonalisation of this Hamiltonian by
means of a Bogoliubov transformation \cite{bogol} the density $n = \langle
\psi^{\dagger}(\vec{x}) \psi(\vec{x}) \rangle$ and the order parameter
$\Delta_0 = V_0 \langle \psi(\vec{x}) \psi(\vec{x}) \rangle$ are easily
calculated, resulting in the equation of state
\begin{equation}
\label{state}
n = \frac{1}{V} \sum_{\vec{k}}
    \left\{ \frac{\epsilon_{\vec{k}} - \mu'}
                 {\hbar \omega_{\vec{k}}}
                               N(\hbar \omega_{\vec{k}})
    + \frac{\epsilon_{\vec{k}} - \mu' - \hbar \omega_{\vec{k}}}
           {2 \hbar \omega_{\vec{k}}} \right\}
\end{equation}
and the BCS gap equation \cite{kleinert,evans}
\begin{eqnarray}
\frac{1}{V_0} + \frac{1}{V} \sum_{\vec{k}}
   \frac{1 + 2N(\hbar \omega_{\vec{k}})}
        {2 \hbar \omega_{\vec{k}}} = 0 \:,   \nonumber
\end{eqnarray}
respectively.

As will become clear in a moment it is important to note that in the dispersion
$\hbar \omega_{\vec{k}} = \sqrt{(\epsilon_{\vec{k}}-\mu')^2 - |\Delta_0|^2}$ of
the Bogoliubov quasiparticles there is a minus sign in front of $|\Delta_0|^2$
instead of the usual plus sign since we are  dealing with paired bosons and not
with paired fermions. Moreover, the gap equation has an ultraviolet divergence
due to the neglect of the momentum dependence of the interaction. However,
anticipating on the fact that $|\Delta_0|$ is at most of
$O(|nT^{2B}(\vec{0},\vec{0};0)|)$ and thus much smaller than $\hbar^2/mr_V^2$,
we find from the Lippmann-Schwinger equation \cite{glockle} for the two-body T
matrix that this divergence is cancelled by a renormalization of $1/V_0$ to
$1/T^{2B}(\vec{0},\vec{0};0)$. Therefore the gap equation becomes
\begin{equation}
\label{gap}
\frac{1}{T^{2B}(\vec{0},\vec{0};0)}
   + \frac{1}{V} \sum_{\vec{k}}
                 \frac{N(\hbar \omega_{\vec{k}})}
                      {\hbar \omega_{\vec{k}}} = 0 \:,
\end{equation}
which is free of divergencies. Together with Eq.\ (\ref{state}) it determines
both $\mu'$ and $|\Delta_0|$ for a fixed density and temperature.

At high temperatures the gap equation has no solution and we must use
$|\Delta_0|=0$. In that case Eq.\ (\ref{state}) reduces to the expected
equation of state for a dilute Bose gas in the normal phase (cf. Eq.\
(\ref{n})). Below the critical temperature $T_{ER}$, determined by a
linearization of Eq.\ (\ref{gap}) which correctly leads to Eq.\ (\ref{ter}),
the order parameter $|\Delta_0|$ becomes nonzero and the gas is in a superfluid
phase with paired atoms. Lowering the temperature, both $|\Delta_0|$ and $\mu'$
increase, but in such a manner that the gap in the dispersion of the Bogoliubov
quasiparticles decreases since otherwise the system would not be able to
accomodate the same number of particles. Evidently, this behavior is possible
due to the above mentioned minus sign in the dispersion relation.

At a second critical temperature $T_{BEC}$ the gap closes and the number of
particles with zero momentum diverges, which signals a Bose-Einstein
condensation. Below that temperature we have $|\Delta_0|=-\mu'$ and $\hbar
\omega_{\vec{k}}$ equals the famous Bogoliubov dispersion
$\sqrt{\epsilon_{\vec{k}}^2 - 2\mu'\epsilon_{\vec{k}}}$ \cite{bogol}.
Furthermore, Eqs.\ (\ref{state}) and (\ref{gap}) are replaced by
\begin{equation}
n = n_0 + \frac{1}{V} \sum_{\vec{k} \neq \vec{0}}
    \left\{ \frac{\epsilon_{\vec{k}} - \mu'}
                 {\hbar \omega_{\vec{k}}}
                               N(\hbar \omega_{\vec{k}})
    + \frac{\epsilon_{\vec{k}} - \mu' - \hbar \omega_{\vec{k}}}
           {2 \hbar \omega_{\vec{k}}} \right\}
\end{equation}
and
\begin{equation}
\frac{1}{T^{2B}(\vec{0},\vec{0};0)} - \frac{n_0}{\mu'}
   + \frac{1}{V} \sum_{\vec{k} \neq \vec{0}}
                 \frac{N(\hbar \omega_{\vec{k}})}
                      {\hbar \omega_{\vec{k}}} = 0 \:,
\end{equation}
determining now $\mu'$ and the condensate density $n_0$ for a fixed density and
temperature. Notice that the mechanism for Bose-Einstein condensation is
identical to the mechanism causing Bose-Einstein condensation in the ideal Bose
gas. In particular, there is no spontaneous breaking of symmetry associated
with the second transition and the order parameter is $n_0$ and not $\langle
\psi(\vec{x},\tau) \rangle$, which is zero also below $T_{BEC}$.

To discuss the question of the mechanical stability of the various phases and
consequently the experimental observability of the two transitions we must
calculate the pressure of the gas. Since the Hamiltonian in Eq.\ (\ref{hamil})
is quadratic this is easily accomplished by evaluating the thermodynamic
potential $\Omega = -pV$ and we find
\begin{eqnarray}
p = \frac{1}{2V} \sum_{\vec{k}}
      \left\{ \epsilon_{\vec{k}} - \mu'
             - \frac{|\Delta_0|^2}{2(\epsilon_{\vec{k}} - \mu')}
             - \hbar \omega_{\vec{k}} \right\}
    + \frac{|\Delta_0|^2}{2T^{2B}(\vec{0},\vec{0};0)}
    + n^2 T^{2B}(\vec{0},\vec{0};0)     \nonumber \\
    - \frac{k_BT}{V} \sum_{\vec{k}}
           ln \left( 1-e^{-\beta \hbar \omega_{\vec{k}}} \right)
                                                             \:,
\end{eqnarray}
where we have again cancelled an ultraviolet divergence in the expression by
renormalizing $1/V_0$ to $1/T^{2B}(\vec{0},\vec{0};0)$. The above expression is
valid for all temperatures if we use $|\Delta_0|=0$ above $T_{ER}$ and
$|\Delta_0|=-\mu'$ below $T_{BEC}$.

In Fig.\ \ref{fig2} we show the behavior of the pressure as a function of
inverse density at a fixed temperature. At high densities the pressure stays
negative, which is unphysical and caused by the fact that the T-matrix
approximation incorporates the hard core of the interatomic potential only in
an effective way. For a dilute gas $nr_C^3 \ll 1$ and this is indeed justified.
At higher densities, however, the hard core becomes of the utmost importance to
avoid the complete collapse of the system and must be treated with more care.
Nevertheless, we can conclude from Fig.\ \ref{fig2} that the gas passes in
thermal equilibrium through a first-order transition from a gaseous phase to a
phase with high density. The dense phase cannot be described accurately within
the framework of the ladder approximation and a more advanced theory, capable
of describing a strongly interacting system, is required to discuss whether it
is liquid or solid and also its possible superfluid properties.

However, as explained in the introduction, for the experimental observability
of the Evans-Rashid transition we only need to answer the question if it can
take place in the (meta)stable region of the phase diagram, where $\partial
p/\partial n >0$. It is not difficult to show (by comparing Eq.\ (\ref{spin})
below with Eq.\ (\ref{ter})) that this is never the case. Therefore, we expect
the following behavior of the atomic system: By increasing the density or
lowering the temperature, the gas will evolve from a stable to a metastable
state until, when $\partial p/\partial n = 0$, a point on the spinodal line is
reached. Increasing the density further the gas is quenched into the unstable
region of the phase diagram and will experience a phase separation by means of
a spinodal decomposition.

\section{CONCLUSIONS AND DISCUSSION}
\label{concl}
In the case of trapped atomic gas samples such a spinodal decomposition should
be easily visible experimentally because it causes an increase in the density
at the center of the trap and consequently leads to a sudden increase in the
number of two-body relaxation and three-body recombination processes that
severely limit the lifetime of the system. Therefore, we present in Fig.\
\ref{fig3} the spinodal line, following from the condition
\begin{equation}
\label{spin}
\frac{1}{T^{2B}(\vec{0},\vec{0};0)}
 + \frac{1}{V} \sum_{\vec{k}}
   \frac{N(\epsilon_{\vec{k}}-\mu')}{\epsilon_{\vec{k}}} = 0 \:.
\end{equation}
We notice that the degeneracy parameter $n\Lambda^3$ is always smaller than
$\zeta(\frac{3}{2}) \simeq 2.612$, implying that it is easier to obtain the
required condition for phase separation than the condition for Bose-Einstein
condensation in the case of an atomic gas with positive scattering length. This
is particularly true for atomic cesium in the $|F=3, M_F=-3 \rangle$ state,
which can have large negative values of the scattering length by tuning the
magnetic field strength. Whether the dynamics of the spinodal decomposition is
also observable is unclear at this moment because it requires a detailed study
of the growth of the liquid or solid phase taking into account the release of
latent heat and the enhanced importance of inelastic collision processes that
lead to decay of the sample.

Summarizing, we have studied the dilute atomic Bose gas with negative
scattering length and have shown that although Bose-Einstein condensation can
in principle also occur in this case, it does not take place in the gas phase.
In contrast, we argue that the gas separates into a normal gas and a (possibly
superfluid) liquid or solid. For magnetically trapped atomic lithium or cesium
the most likely scenario is phase separation into a gas and a solid since both
the thermal energy as well as the energy due to zero-point motion are much
smaller than the depth of the potential well. However, a more elaborate
discussion is necessary to confirm this conjecture. In any case the separation
proceeds by means of a spinodal decomposition and gives a clear experimental
signal if one monitors the decay of the atomic density.

\section*{ACKNOWLEDGEMENTS}
It is a pleasure to acknowledge various helpful discussions with Michel
Bijlsma, Ard-Jan Moerdijk, Eite Tiesinga and Randy Hulet. I also thank Tony
Leggett for pointing out reference \cite{evans} to me.

\begin{figure}
\caption{Qualitative picture of the relationship between pressure
         and inverse density at a fixed and sufficiently low
         temperature for 1) atomic lithium or cesium
         and 2) atomic hydrogen. Note that
         possible cusps in the curves due to symmetry-breaking
         phase transitions are not shown here.
         \label{fig1}}
\end{figure}

\begin{figure}
\caption{Pressure as a function of inverse density for a fixed
         temperature such that $|a/\Lambda|=10^{-1}$. The inset
         shows the cusp in the pressure caused by the
         Evans-Rashid transition.
         \label{fig2}}
\end{figure}

\begin{figure}
\caption{Spinodal line in the ($n\Lambda^3, na^3$) plane,
         separating the (meta)stable and unstable regions of
         the phase diagram. The dashed line corresponds to the
         condition for Bose-Einstein condensation in a Bose gas
         with repulsive interactions.
         \label{fig3}}
\end{figure}


\begin{references}
\bibitem{foot} At low densities the behavior of curve 1 is
               determined in Sec.\ \ref{stable}. Its behavior at
               high densities is based on the following two
               observations. First, phase transitions
               that possibly take place along the
               isotherm can only lead to cusps in the curve and
               not to discontinuities or divergencies. For
               discontinuous (first order) phase transitions this
               is true by definition and for continuous phase
               transitions it follows from the renormalization
               group analysis, which shows that the
               singular part of the thermodynamic potential
               behaves as $V\Omega_{\pm}/\xi^d$, where $d$ is
               the dimension and the correlation length $\xi$
               diverges as $(n_c/(n-n_c))^{\nu}$ near the
               critical density $n_c$ \cite{huang}. Second, due
               to the repulsive core of the interatomic
               potential the pressure must always become positive
               at a density of $O(r_c^{-3})$ and diverge
               for $n \rightarrow \infty$.
\end{references}
\end{document}